\documentstyle[prl,twocolumn,aps]{revtex}
\input epsf
\begin{document} \draft 
\twocolumn[
\title{Numerical Evidence for Multiplicative Logarithmic Corrections 
from Marginal Operators}
\author{Sebastian Eggert\cite{email}}
\address{Theoretical Physics, Chalmers University of Technology
 and G\"oteborg University, 41296 G\"oteborg, Sweden}
  \date{\today} \maketitle 
\begin{abstract}
\widetext\leftskip=0.10753\textwidth \rightskip\leftskip
Field theory calculations predict multiplicative logarithmic corrections
 to correlation functions
from marginally irrelevant operators. However, for the numerically most
 suitable model -- the spin-1/2 chain -- these corrections have been 
 controversial. In this paper, 
the spin-spin correlation function of the 
antiferromagnetic spin-1/2 chain
is calculated numerically in the presence of a next nearest neighbor 
coupling $J_2$ for chains of up to 32 sites. By varying the 
coupling strength $J_2$ we can control the effect of the marginal operator,
and our results unambiguously confirm the field theory predictions. 
The critical value at which the marginal operator vanishes has been determined
to be at $J_2/J = 0.241167 \pm 0.000005$.
\end{abstract}
] \narrowtext
The spin-1/2 chain has attracted much attention in theoretical physics
since the early days of quantum mechanics as a simple model for many-body
effects.  However, correlation functions in the antiferromagnetic chain
could only be predicted with the help of field theory 
calculations\cite{peschel},
and it was only a few years ago when a prediction of a {\it multiplicative}
logarithmic correction to the asymptotic spin-spin correlation functions 
was made coming from a marginally irrelevant operator in the
field theory description\cite{affleck,schulz}.  
At first sight this result seems surprising, since it indicates that
the effect of a marginally irrelevant operator effectively {\it increases} 
in the long distance limit, while on the other hand its effect
on other quantities 
(energy spectrum\cite{affleck}, susceptibility\cite{taka}) vanishes 
as length \hbox{$L\to \infty$} and  temperature \hbox{$T \to 0$}. 

However, the underlying calculation which predicted the multiplicative 
correction from a marginal operator is quite general\cite{amit} 
and of great importance in other models as well (e.g. the $\phi^4$-model).
Since the spin-1/2 chain is accessible by a number of numerical methods 
it represents an ideal testing model  for this important result.
Therefore, considerable effort has been made to verify the logarithmic
 corrections\cite{hallberg,scalap,studies}, but often different 
or no multiplicative corrections were found.

In the present paper we choose to numerically diagonalize the model 
Hamiltonian  of up to 32 sites exactly.  While this method is limited by
 relatively small system sizes, it allows us to introduce  a next nearest
neighbor coupling $J_2$ quite easily and, moreover,
 the results are very accurate.
It is known that by adjusting $J_2$ we can change the bare coupling
strength of the marginal operator and even set it to zero without destroying
the validity of the field theory continuum limit\cite{julien}.  
Since we are able to probe the system with a reduced
marginal coupling, the range of validity for
the perturbative field theory calculations is increased
exponentially.  Moreover, we will use a novel approach to analyze the
data which does not depend on any extrapolation to an infinite system 
size for the field theory calculations to be valid, so that 32 sites 
turn out to be very sufficient to confirm the effect unambiguously. 

Our model Hamiltonian is the antiferromagnetic spin-1/2 chain $(J > 0)$
with a next nearest neighbor coupling $J_2$
\begin{equation}
H \ = \ J \sum_x  {\bf S}_x\cdot {\bf S}_{x+1} \ + \ 
J_2 \sum_x  {\bf S}_x \cdot {\bf S}_{x+2}.\label{H}
\end{equation}
In the long-wavelength, low-energy limit this model is well understood
in terms of the Wess-Zumino-Witten (WZW) non-linear $\sigma$-model with
a topological coupling constant $k =1$\cite{WZW}.  In the continuum limit,
the spin-1/2 operator ${\bf S}_x$ at each lattice 
site $x$  can then be expressed in 
terms of a WZW SU(2) matrix ${g_{\alpha}^{\beta}}$ and (related) chiral SU(2) 
currents ${\bf J}_{L,R}$
\begin{equation}
{\bf S}_x \ = \  \left( {\bf J}_L + {\bf J}_R \right) \ + \  
{\rm const.}i (-1)^x {\rm tr}[g  \mbox{\boldmath{$\sigma$}}].
\label{spin}
\end{equation}

Strictly speaking this theory is only valid up to correcting
higher order operators in the Hamiltonian
which we have neglected so far.  In particular, we want to consider the
effect of the leading irrelevant marginal operator which can be written
in terms of the current operators with some coupling constant $\lambda$
\begin{equation}
\delta {\cal H} = - 2 \pi \lambda {\bf J}_L \cdot {\bf J}_R.
\label{marginal}
\end{equation}
For periodic boundary condition the next ``leading''  irrelevant operator
in the Hamiltonian has scaling dimension 4, which can  
be neglected  even for modest lengths of order \hbox{$L \agt 10$} and 
temperatures of order \hbox{$T\alt 0.1J$}\cite{eggert}.  

The effect of the marginal operator on the uniform 
(i.e. current-current) part of the spin-spin correlation function 
is well understood in terms of a simple additive logarithmic
correction.  This leads to interesting predictions for the asymptotic
behavior of the susceptibility\cite{taka}, 
which were well confirmed by Bethe ansatz
and also experimental results\cite{expsusc}.
The situation is quite different, however, for the alternating part of the
spin-spin correlation function $G(r)$ which is given in terms of 
the WZW field ${g}$
\begin{equation}
G(r) \propto < {\rm tr}[ g   \mbox{\boldmath{$\sigma$}}](0)\cdot
{\rm tr}[g   \mbox{\boldmath{$\sigma$}}](r)>.\label{corrfcn}
\end{equation}
For the primary field $g$ 
we have to take into account the effect of the marginal operator on the
anomalous scaling dimension $\gamma(\lambda)$ of ${\rm tr}
[ g  \mbox{\boldmath{$\sigma$}}]$
as well (while the scaling dimension of
the currents is fixed).  
The correlation function then also depends on the marginal
coupling constant $\lambda$, and $G(r,\lambda)$ obeys the  
renormalization equation 
\begin{equation}
\left[ \frac{\partial}{\partial \ln r} + \beta(\lambda) 
\frac{\partial}{\partial \lambda} + 2 \gamma(\lambda)\right] G(r,\lambda) = 0
\label{renorm}
\end{equation}
where $\beta(\lambda)$ is the beta function of the marginal operator 
$\delta \cal H$. 
This equation (\ref{renorm}) is then solved by
introducing a scale dependent coupling constant $\lambda(r)$ which obeys
\begin{equation}
\frac{\partial \lambda}{\partial \ln r} = \beta(\lambda) \label{beta}
\end{equation}
and writing (up to a prefactor which may contain additive corrections
in $\lambda$)
\begin{eqnarray}
G(r) & \propto & \exp \left[-2\int^{\ln r}_{\ln r_1} \gamma(x)d(\ln x)
\right]\nonumber \\
& = & \exp \left[-2\int^{\lambda(r)}_{\lambda_0}
\frac{\gamma(\lambda)}{\beta(\lambda)} d \lambda \right],  \label{solution}
\end{eqnarray}
where $r_1$ is an  ultraviolet cut-off of the order of the lattice spacing 
and $\lambda_0$ is the unrenormalized (bare) 
coupling strength at that energy-scale.
The beta-function and the anomalous dimension are known as perturbative
expansions in $\lambda$\cite{affleck}
\begin{eqnarray}
\beta(\lambda) & = & - \lambda^2 + {\cal O}(\lambda^3)
\nonumber \\
\gamma(\lambda) & = & \frac{1}{2} - \frac{\lambda}{4}
+ {\cal O}(\lambda^2).\label{expand}
\end{eqnarray}  
Integrating equation (\ref{beta}), we obtain an expression for  $\lambda$ 
up to ${\cal O}[\ln(\ln r)/(\ln r)^2]$
\begin{equation}
\lambda(r) \approx \frac{\lambda_0}{1+ \lambda_0 \ln(r/r_1)}
= \frac{1}{\ln(r/r_0)}, \label{lambda}
\end{equation}
and together with equation (\ref{solution}) we can determine the
correlation function up to higher orders in $\lambda$ 
\begin{equation}
G(r) \propto \frac{\left(\lambda/\lambda_0\right)^{-1/2}}{r}
\left[1 + {\cal O}(\lambda)\right] \approx
\frac{\sqrt{\lambda_0 \ln r/r_0}}{r}
\label{expansion}
\end{equation}
From equation (\ref{lambda}) we see that the constant $r_0$ is simply given by
\begin{equation} r_0 = r_1 \exp(-1/\lambda_0), \label{r_0}\end{equation}  
In the asymptotic limit we find $G \propto \sqrt{\ln r}/r$ as already predicted
in reference \onlinecite{affleck} unless $\lambda_0$ is infinitesimally small.
Therefore, very close to the critical point the predicted 
asymptotic limit will never be reached although the perturbative expansions 
in equation (\ref{expand}) are very accurate.  
Since $r_0$ is a non-universal constant, 
it is essential to keep the full expression in equation (\ref{expansion})
to test that result (as opposed to just assuming the asymptotic limit).

It is important to note that equation
(\ref{renorm}) is only valid if $r$ is the only length scale in the system, 
but all previous studies assumed that this would require the infinite
length limit \hbox{$L\to \infty$} which is numerically very difficult
to achieve.  However, it is perfectly feasible
to study the system at finite, but varying lengths $L$ and
keeping the ratio $R= r/L$ fixed.  Since $r= R L$ 
is no longer independent of $L,$
there is only  one variable parameter in the problem and therefore
expression (\ref{expansion}) holds for $G(r=RL)$
(for each $R$ separately, i.e.~the 
overall proportionality constant will depend on the choice of $R$). 
To test the field theory predictions it is therefore much easier and more
accurate to select the fixed ratio to be $R=1/2,$ and in what follows we
will hence set  \begin{equation} r \equiv L/2. \end{equation}
It is then straightforward to determine the correlation function 
$G(r=L/2)$ numerically 
for different lengths $L \le 32$ and coupling constants $J_2$ (as opposed to
selecting $R=0$ and having to choose some controversial\cite{hallberg,scalap} 
extrapolation to the \hbox{$L \to \infty$} limit as in all previous studies).

The correlation functions of the ground state
were found by exact diagonalization using 
the modified Lanczos method on periodic spin chains of up to 32 sites
(using ca.~4.7 million basis states). 
Table \ref{results} shows the results for the spin-spin correlation
functions $<S_z(0) S_z(r)>$  
for different distances $r=L/2$ and for various values of the next nearest
neighbor coupling $J_2$.

By adjusting $J_2$ we can change the bare coupling $\lambda_0$ 
and even set it to zero at some critical value $J_2^{\rm crit}$\cite{julien}.
To lowest order, $\lambda_0$ increases linearly with the difference from the
critical point \hbox{$\Delta J_2 \equiv J_2^{\rm crit} - J_2$.}
The critical value $J_2^{\rm crit}$
was determined to be \begin{equation}
J^{\rm crit}_2 \approx 0.241167\pm 0.000005J, \label{J_crit} \end{equation} 
which we believe to be
the most accurate estimate to date.  This result 
was obtained by examining the energy difference $\Delta$
between the first two excited states of total spin $S=0$ and $S=1$ 
in the periodic chain spectrum 
as a function of length $L$ and $J_2$.   Those two states are nearly 
degenerate and their energy difference is only due to the marginal operator
and higher order terms in the field theory Hamiltonian.  
\twocolumn[
\begin{table}
\widetext
\begin{tabular}{ccc|rrrrrr}
&$r=L/2$ && $J_2= -0.25J$ & $J_2= 0$& $J_2= 0.1J$&
$J_2= 0.2J$& $J_2= J_2^{\rm crit}$ \\
\hline
&  4&& 0.0572167& 0.0497077& 0.0459760& 0.0417541& 0.0398672&\\
&  5&&-0.0545609&-0.0469513&-0.0433261&-0.0390852&-0.0370562&\\
&  6&& 0.0423313& 0.0356626& 0.0325190& 0.0288716& 0.0271334&\\
&  7&&-0.0403820&-0.0340938&-0.0311410&-0.0276488&-0.0259367&\\
&  8&& 0.0336161& 0.0279325& 0.0252780& 0.0221537& 0.0206302&\\
&  9&&-0.0322537&-0.0269130&-0.0244137&-0.0214326&-0.0199523&\\
& 10&& 0.0279427& 0.0230233& 0.0207281& 0.0179995& 0.0166504&\\
& 11&&-0.0269492&-0.0223039&-0.0201300&-0.0175197&-0.0162116&\\
& 12&& 0.0239555& 0.0196206& 0.0175960& 0.0151709& 0.0139598&\\
& 13&&-0.0232004&-0.0190837&-0.0171551&-0.0148271&-0.0136519&\\
& 14&& 0.0209966& 0.0171185& 0.0153044& 0.0131186& 0.0120183&\\
& 15&&-0.0204030&-0.0167013&-0.0149649&-0.0128595&-0.0117902&\\
& 16&& 0.0187105& 0.0151987& 0.0135530& 0.0115605& 0.0105511&
\end{tabular}
\caption{The spin-spin correlation function $<S_z(0) S_z(r)>$
for various values of $r = L/2$ and $J_2$.  In some cases the last digit is 
uncertain.} \label{results}
\end{table}] \narrowtext
We find
the critical point  by determining the value of $J_2$ at which the energy
difference $\Delta$ is exactly proportional to $1/L^3$, because at that 
point the correction $\Delta$ comes only from the higher order terms and
therefore the marginal operator is absent\cite{eggert}.  
This approach proves to be more 
accurate than the conventional method of extrapolating
the value of $J_2(\Delta=0)$ as $L\to \infty$ (which of course gives the 
same value, although with less accuracy\cite{j2}).

As we approach $J_2^{\rm crit}$ 
 the value of the effective 
length scale $r_0$ becomes exponentially small 
$r_0 = r_1 \exp(-1/\lambda_0),$
and the expansions
in $1/\ln(r/r_0)$ in equations (\ref{lambda}) and (\ref{expansion})
give very accurate results even for moderate lengths.  However, we require the 
marginal operator to remain the {\it leading} correction (compared
to the $1/L^2$ corrections from higher order operators), so it is not always
useful to examine the system arbitrarily close to $J_2^{\rm crit}$.

\begin{figure}
\epsfxsize=3.3in \epsfbox{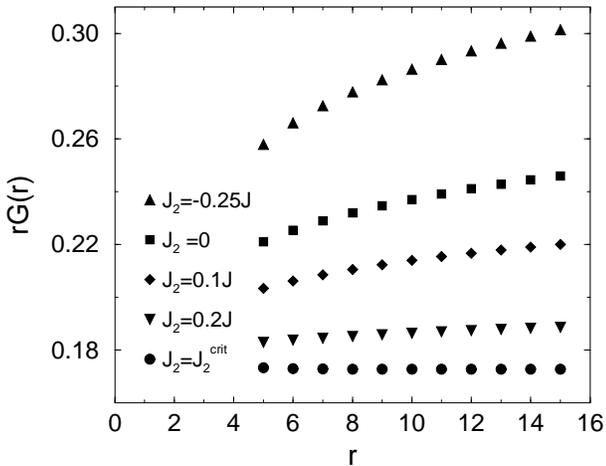}
\caption{The multiplicative correction $r G(r)$ as a function of $r$ for
various values of $J_2$ according to the numerical data in 
table \protect{\ref{results}}.}
\label{all-data}
\end{figure}
\begin{figure}
\epsfxsize=3.3in \epsfbox{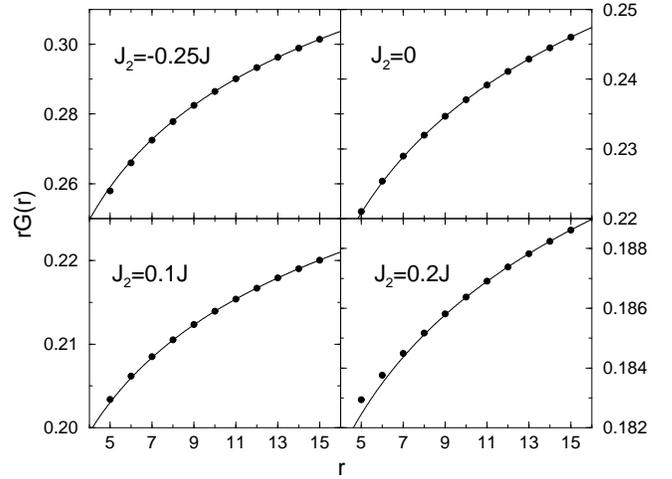}
\caption{The fitting curves to $r G(r)$ according to equation 
(\protect{\ref{fitting-form}}) with the parameters given in
table \protect{\ref{parameters}}.}
\label{fit}
\end{figure}
To extract the multiplicative correction 
we perform a cubic spline interpolation of \hbox{$r<S_z(0) S_z(r)>$}  
from the data
in table \ref{results} for even and odd $r$ separately.  
The difference between the two curves then gives the multiplicative
correction $r G(r)$ to the alternating part of the correlation function
as shown in figure (\ref{all-data}) for different values of $J_2$.

Figure (\ref{fit}) shows the excellent asymptotic fit to the predicted form of 
the multiplicative corrections
\begin{equation} r G(r) =  \sqrt{a\lambda_0\ln r/r_0} 
\label{fitting-form}\end{equation}
with the fitting parameters given in table
\ref{parameters} where $a$ is a normalization constant which
must be independent of $\lambda_0$.
In all cases the data approaches the asymptotic curve 
very quickly. The small deviations are positive for small values of
 $\Delta J_2 \equiv J_2^{\rm crit}-J_2$ and negative for larger 
$\Delta J_2$,  
which indicates a partial cancellation of higher order terms with
varying relative magnitude as $\lambda_0$ is changed (this explains why
the first order calculation already gives such good results).
\vbox{\begin{table}
\begin{center}
\begin{tabular}{ccccc}
&$J_2/J$ & $\ln r_0$ & $a \lambda_0$ & \\
\hline
&-0.25 & -1.4976 & 0.021601 &\\
& -0.1 &  -2.1683     & 0.014864 &\\
&0 & -2.9501& 0.010696&\\
& 0.05 & -3.6111 & 0.008626 &\\
&0.1 &-4.6883 & 0.006546 &\\
& 0.15 & -6.9080 & 0.004383 &\\
&0.2 & -14.393 & 0.002080& 
\end{tabular}
\end{center}
\caption{The fitting parameters for according 
to equation (\protect{\ref{fitting-form}}) for selected values of $J_2$.}
\label{parameters}
\end{table}}

For the  special $J_2 = 0$ case we find general agreement
with reference \onlinecite{hallberg} where a similar fit was made, however with
much larger deviations, and -- more importantly -- their data does not approach
the predicted form asymptotically as it must.
(In that reference an extrapolation scheme with additional adjustable 
parameters was used as well as a less accurate numerical method.)

\begin{figure}
\epsfxsize=3.3in \epsfbox{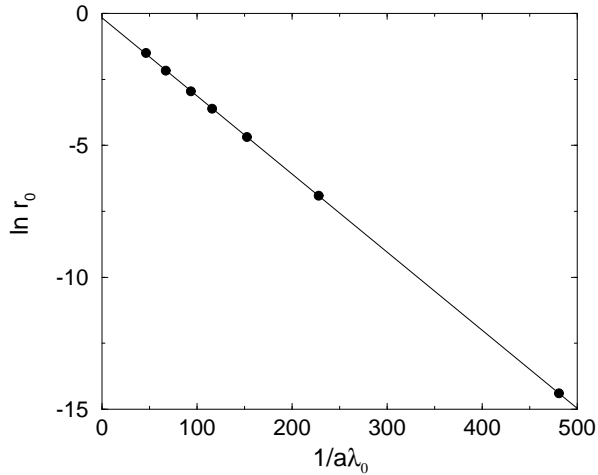}
\caption{The logarithm $\ln r_0$ vs. $1/a \lambda_0$  from table
\protect{\ref{parameters}} shows an excellent fit to  
$\ln r_0 = \ln r_1 - 1/\lambda_0$ with $r_1 = 0.85$ and $a = 0.0296$, 
confirming equation (\protect{\ref{relation}}).}
\label{relate}
\end{figure}
As the critical point is approached, we require simultaneously 
$r_0 \to 0$ and $\lambda_0 \to 0$ in such a way that the correlation 
function $G(r)$ remains finite.  
In particular we know from equation (\ref{r_0}) 
that $r_0$ and $\lambda_0$ are related 
(to lowest order the correlation function $G(r)$ is independent of the 
size of the marginal coupling  at the ultraviolet cutoff $r_1$) i.e.
for all $J_2 \leq J_2^{\rm crit}$ we must have:
\begin{equation}
\lambda_0 \ln r_1/r_0 = 1 \label{relation}
\end{equation}
As can be seen in figure (\ref{relate}) this relation holds very
accurately for the parameters in table \ref{parameters} 
for all values of $J_2$ with the constant in
equation (\ref{fitting-form}) given by 
$a = 0.0296$.  Equation (\ref{relation}) also fixes the ultraviolet
cutoff $r_1 = 0.85$ at which the bare coupling constant is defined.
Because  $a$ and $r_1$ are fixed, $\lambda_0$ and $r_0$ are related by 
equation (\ref{relation}), and there remains really only one free parameter 
for each of the curves in figure (\ref{fit}), which makes the quality of the 
fits even more impressive. 

\begin{figure}
\epsfxsize=3.3in \epsfbox{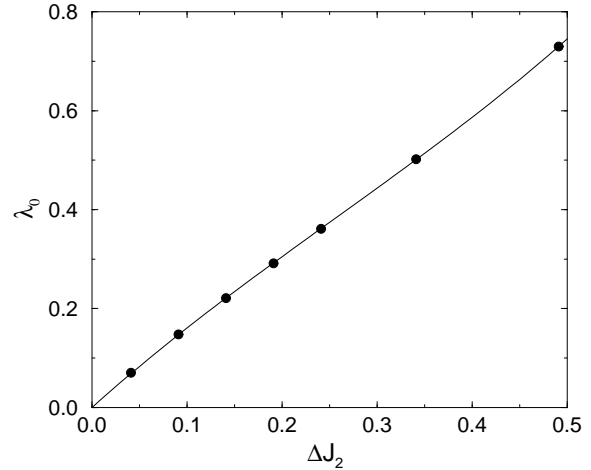}
\caption{The parameter $\lambda_0$ vs. $\Delta J_2 \equiv J_2^{\rm crit}-J_2$ 
from table \protect{\ref{parameters}} is fitted
to  equation (\protect{\ref{c}}).}
\label{bare}
\end{figure}
Moreover, analyzing the fitting parameters in table \ref{parameters}, 
we can determine the functional behavior of the
bare coupling $\lambda_0$ in form of a series expansion in the
microscopic parameter $\Delta J_2 = J_2^{\rm crit} - J_2$ 

\begin{equation}
\lambda_0 = c_1 \Delta J_2 + c_2 \Delta J_2^2 + c_3 \Delta J_2^3, \label{c}
\end{equation}
As shown in figure (\ref{bare}), this relation (\ref{c})
seems to hold even for
relatively large values of $\Delta J$ with 
$c_1 \approx 1.723/J, \ c_2 \approx -1.35/J^2,$ and $c_3 \approx 1.76/J^3$.  
Therefore, once the constants $a, \ r_1,$ and $c_i$ are known, the shape 
and overall magnitude of $G(r)$ can be predicted accurately 
for a large range of values $J_2$ and $r \agt 5$, which is a clear 
success of the field theory calculation. 

In conclusion, we have unambiguously shown that the field theory 
calculations for multiplicative corrections to the correlation
functions in the spin-1/2 chain are valid.  The constants $a, \ r_1,$ and $c_i$
were determined, and we were able to improve the estimate of $J_2^{\rm crit}$.
The basic result  carries over to
other models with marginal operators (1-D Hubbard model, other half-integer
spin chains, $\phi^4$-models) where we also expect that the corresponding
field theory calculation gives accurate predictions to multiplicative
corrections to correlation functions. 

\begin{acknowledgements}
The author is very grateful for a number of extremely helpful 
communications with Ian Affleck.  Special thanks go to 
Henrik Johannesson and Stefan Rommer for comments on the manuscript.  
This research was
supported in part by the Swedish Natural Science Research Council.
\end{acknowledgements}


\begin{references} 
\bibitem{email} E-mail address: eggert@fy.chalmers.se
\bibitem{peschel} A.~Luther, I.~Peschel, Phys. Rev.
{\bf B12}, 3908 (1975). 
\bibitem{affleck} I.~Affleck, D.
Gepner, H.J. Schulz, T. Ziman, J. Phys.  {\bf A22}, 511 (1989).
\bibitem{schulz} T.~Giamarchi, H.J.~Schulz, Phys. Rev. {\bf B39}, 4620 (1989);
R.R.~Singh, M.F.~Fischer, R.~Shankar, Phys. Rev. {\bf B39}, 2562 (1989).
\bibitem{taka} S.~Eggert, I.~Affleck, M.~Takahashi, 
Phys. Rev. Lett. {\bf 73}, 332 (1994).
\bibitem{amit} For a review of this ``text-book calculation'' see
D.J.~Amit {\it Field Theory, the Renormalization Group, and Critical 
Phenomena} (World Scientific, Singapore, 1989).
\bibitem{hallberg} K.A.~Hallberg, P.~Horsch, G.~Mart\'inez, Phys. Rev.
{\bf B52}, R719, (1995).
\bibitem{scalap} A.W.~Sandvik, D.J.~Scalapino, Phys. Rev. {\bf B47}, 
12333 (1993).
\bibitem{studies} K.~Kubo, T.A.~Kaplan, J. Borysowicz, Phys. Rev. {\bf B38},
11550 (1988); S.~Liang, Phys. Rev. Lett. {\bf 64}, 1597 (1990);
H.Q.~Lin, D.K. Campbell, J. Appl. Phys. {\bf 69}, 5947 (1991).
\bibitem{julien} R.~Julien, F.D.M.~Haldane, Bull. Am. Phys. Soc. {\bf 28}, 34
(1983).
\bibitem{WZW} For a review and a list of original references see
I.~Affleck, {\it Fields,
Strings and Critical Phenomena}  (ed. E. Br\'ezin, J. Zinn-Justin),
(North-Holland, Amsterdam 1990), p.563. 
\bibitem{eggert}S.~Eggert, I.~Affleck, Phys.\
Rev.\ {\bf B46}, 10866, (1992).
\bibitem{expsusc} T. Ami, M.K. Crawford, R.L. Harlow, Z.R. Wang,
D.C. Johnston, Q. Huang, R.W. Erwin, Phys. Rev. {\bf B51}, 5994 (1995);
S.~Eggert, Phys. Rev. {\bf B53}, 5116 (1996).
\bibitem{j2} K.~Okamoto, K.~Nomura, Phys. Lett. {\bf A169}, 433 (1992);
G.~Castilla, S.~Chakravarty, V.J.~Emery, Phys. Rev. Lett. 
{\bf 75}, 1823 (1995).
\end{references}
\end{document}